\documentclass[printer]{aa}

\bibpunct{(}{)}{;}{a}{}{,}

\usepackage{lscape}
\usepackage{graphicx}
\usepackage[varg]{txfonts}
\newcommand{\hii}{H\textsc{ii}}
\def\ks{km s$^{-1}$}
\def\d{$^\circ$}
\def\m{$^\prime$}
\def\s{$^{\prime\prime}$}
\def\hh{$^{\mathrm h}$}
\def\mm{$^{\mathrm m}$}
\def\ss{$^{\mathrm s}$}
\def\cm3{cm$^{-3}$}

\def\2{$^{12}$CO}
\def\3{$^{13}$CO}
\def\8{C$^{18}$O}
\def\msol{M$_\odot$}

\def\cm2{cm$^{-2}$}

\def\g29{G29.96$-$0.02}

\begin{document}

\title{Studying star-forming processes at core and clump scales: the case of the young stellar object G29.862$-$0.0044 }
\author {M.B. Areal \inst{1}
\and S. Paron \inst{1}
\and C. Fariña \inst{2,3}
\and M.E. Ortega \inst{1}
\and M. Celis Peña \inst{1}
\and M. Rubio \inst{4}
}

\institute{CONICET - Universidad de Buenos Aires. Instituto de Astronom\'{\i}a y F\'{\i}sica del Espacio
             CC 67, Suc. 28, 1428 Buenos Aires, Argentina
\and Isaac Newton Group of Telescopes, E38700, La Palma, Spain
\and Instituto de Astrof\'{\i}sica de Canarias (IAC) and Universidad de La Laguna, Dpto. Astrof\'{\i}sica, Spain
\and Departamento de Astronom\'{\i}a, Universidad de Chile, Casilla 36-D, Santiago, Chile
}

\offprints{M.B. Areal}

   \date{Received <date>; Accepted <date>}

\abstract{}{To advance in our knowledge about star formation, besides the statistical studies and large surveys of 
young stellar objects (YSOs), it is important to do detailed studies towards particular objects. Given that massive molecular clumps 
fragment into cores where star formation takes place, this kind of studies should be done at different spatial scales.}
{Using near-IR data obtained with NIRI at the Gemini-North telescope, 
data of the complex molecular species CH$_{3}$OCHO and CH$_{3}$CN obtained from the Atacama Large Millimeter Array (ALMA) database, observations of HCN, HNC, HCO$^{+}$, and C$_{2}$H carried out with 
the Atacama Submillimeter Telescope Experiment (ASTE), and CO data from public surveys observed with the James Clerck Maxwell Telescope (JCMT), we perform a deep study of the YSO G29.862$-$0.0044 (YSO-G29) at core and clump spatial scales.}
{The near-IR emission shows that YSO-G29 is composed by two nebulosities separated by a dark lane, suggesting a scenario consistent with a typical disk-jets system, but in this case, highly asymmetric. The northern nebulosity is open, diffuse and it is divided into two branches, while the southern one is smaller and sharper. They are likely produced by the scattered light in cavities carved out by jets or winds on an infalling envelope of material, which also present line emission of H$_{2}$ S(1) 1--0 and 2--1, and [FeII]. The presence of the complex molecular species observed with ALMA confirms that we are mapping a hot molecular core. The CH$_3$CN emission concentrates at the position of the dark lane and it appears slightly elongated from southwest to northeast in agreement with the inclination of the system as observed at near-IR. The morphology of the CH$_3$OCHO emission is more complex and extends along
some filaments and concentrates in knots and clumps, mainly southwards the dark-lane, suggesting that the southern jet is encountering a dense region. The northern jet can flow more freely, generating the more extended features as seen at near-IR. This is in agreement with the red-shifted molecular outflow traced by the \2 J=3--2 line extending towards the northwest and the lack of a blue-shifted outflow. 
This configuration can be explained by considering that G29-YSO is located at the furthest edge of the molecular clump along the line of sight, which is consistent with the position of the source in the cloud mapped in the \8 J=3--2 line. The detection of HCN, HNC, HCO$^{+}$, and C$_{2}$H allowed us to characterize the dense gas at clump scales, yielding results that are in agreement with the presence of a high-mass protostellar object.}
{}

\titlerunning{Star-forming processes at core and clump scales}
\authorrunning{M.B. Areal et al.}


\maketitle

\section{Introduction}
\label{intro}

It is known that massive stars form deeply embedded in cores of giant molecular clouds,
places with very high visual absorption due to the presence of abundant interstellar dust. 
They form on relatively short time-scales, at distances greater than
the nearer examples of their low-mass counterparts \citep{hill05}. Additionally, given
that massive stars tend to form in clusters, the regions in which they form are very confused.
These issues make it difficult to obtain useful observational data towards individual massive young
stellar objects (MYSOs), and hence, our knowledge about the physics of the massive star formation 
is far from being complete.

MYSOs produce massive molecular outflows \citep{kurtz00,arce07}
which contribute to the removal of excess angular momentum from accreted matter and to disperse infalling
circumstellar envelopes \citep{reip01,prei03}. The study of molecular outflows and accretion processes is useful to improve our
understanding of the formation of stars of all masses, in particular of high-mass stars \citep{yang18}.
Thus, even though the molecular outflows can be frequently confused with the molecular gas of the clump
in which the MYSOs are embedded and with gas associated with other nearby sources, it is worth making efforts in 
studying them together with the driven sources. Hence, besides the large surveys and statistical studies about YSOs and 
massive outflows, it is also necessary to do detailed studies towards particular objects. Moreover, given that 
it is known that massive clumps fragment into cores where star formation takes place \citep{motte18}, this kind of studies 
should be done at different spatial scales, characterizing the physical and chemical properties of the star-forming regions at clump and cores scales (see \citealt{sch19} and previous papers of the series).

Nowadays there are very useful surveys of MYSOs and molecular outflows generated from sources of the Red MSX Source (RMS) 
survey database \citep{lumsden13}. For instance, \citet{maud15a}, based on \8 data, analyze a large sample of massive star-forming cores
and then, \citet{maud15b} study associated molecular outflows using \2 and \3 data. 
More recently, \citet{yang18} present the largest survey of outflows within the Galactic plane using \3 and \8 data.  
These surveys provide large and homogeneous samples of sources that allow us to select particular MYSOs to perform dedicated observations for deeper and more detailed studies on individual objects. 

In this work we focus on YSO G29.862$-$0.044 (hereafter YSO-G29), which 
is embedded in the southern portion
of a molecular cloud located at a distance of about 6.2 kpc, related to the star-forming region G29.96$-$0.02 (W43-South, \citealt{carl13}).  
This YSO is likely embedded in a dense cold dust clump traced by both 870 $\mu$m and 1.1 mm emissions (\citealt{urqu14} and \citealt{roso10}, respectively).
YSO-G29 presents NH$_{3}$ and methanol maser emission at v$_{\rm LSR} \sim$ 101 \ks~\citep{urqu11,pesta05}, 
and according to \citet{li16}, who analyze \2 and \3 J=1--0 data, only a red molecular outflow is observed, which has an estimated  mass of 14 \msol.  
\citet{yang18}, based on an analysis of the \3 and \8 J=3--2 emission, report
velocity ranges for the blue and red wings in the \3 spectra of 94.7--100.2 and 103.2--106.2 \ks, respectively, suggesting the presence of 
blue and red outflows. However, they do not determine any outflow parameter for this source, probably due to the confusion with the ambient molecular
gas and/or with the cold dust clump in which the YSO is likely embedded.

We present new high-resolution near-IR data obtained with NIRI at the Gemini-North telescope and an analysis of Atacama Large Millimeter Array (ALMA) 
data towards YSO-G29 to study in detail the source and its surroundings at core scales. Additionally, observations of HCN, HNC, HCO$^{+}$, and C$_{2}$H obtained with the Atacama Submillimeter Telescope Experiment (ASTE), and an analysis
of CO public data are presented to study the molecular environment of YSO-G29 at clump scales.

\section{Data and data reduction}

\subsection{Near-IR data}

The images were acquired with the Near InfraRed Imager and Spectrometer (NIRI; \citealt{hodapp03}) at the Gemini-North 8.2-m telescope.
The observations were carried out on July 2017 in queue mode (Band-1 Program GN-2017B-Q-25).
NIRI was used with the f/6 camera that provides a plate scale of 0\farcs117 pix$^{-1}$ in a field of view of 120\s$\times$120\s.

The extended source in the near-IR occupies a large fraction of the field and the background exhibits nebular emission. Therefore, in addition to the dithering pattern on-source, it was also necessary to perform off-source observations in each filter for sky subtraction.

The NIRI data was reduced with DRAGONS, the Gemini's new Python-base data reduction platform (version 2.1.0) and Theli  \citep{sch13,erben05}. The absolute astrometric solution was made using targets in the field of the 2MASS 6X Point Source Working Database Catalog \citep{cutri}.

Table\,\ref{obs} lists the filters used with their central wavelength and width, the effective spatial resolution measured as the average FWHM
of point sources in the final stacked images, and the effective exposure times of the final images. Note that all images presented in this paper were normalized to 1 sec. No flux calibration was made.

The {\it H}-cont image was used to subtract the continuum of [FeII] and the {\it K}-cont one to subtract the continuum of
H$_{2}$ 1-0 S(1), H$_{2}$ 2-1 S(1), and CO 2-0 (bh). For the continuum subtraction the images were convolved with a Gaussian kernel to achieve a similar PSF for the point sources at the central area of both images.
In this process the effective resolution of both images was degraded to a similar value. After convolution the images were scaled to account
for the differences in filter width and throughput, and other effects derived from possibly variable observing conditions. The scale factors were derived from aperture photometry of point sources in the central part of the field.

\begin{table}
\caption{Near-IR bands observed with NIRI at Gemini-North.}
\label{obs}
\tiny
\centering
\begin{tabular}{lcccc}
\hline
\hline
\noalign{\smallskip}
Filter      & $\lambda_c $ & Width   & Eff. Resol.  & Exp. Time  \\
            & ($\mu$m)      & ($\mu$m) & (arcsec)        & (sec)      \\
\hline
\noalign{\smallskip}
\multicolumn{5}{c}{ {\it Broad-bands}  }\\
\hline
\noalign{\smallskip}
{\it J}           &   1.25       &   0.18  &  0.7         &  160    \\
\noalign{\smallskip}
{\it H}           &   1.65       &   0.29  &  0.5         &   18     \\
\noalign{\smallskip}
{\it Ks}          &   2.15       &   0.35  &  0.4        &   3.25\\
\hline
\noalign{\smallskip}
\multicolumn{5}{c}{ {\it Narrow-bands}  }\\
\hline
\noalign{\smallskip}
{\it H}-cont     & 1.570        &  0.0236 &  0.7          & 3060    \\
\noalign{\smallskip}
[FeII]      & 1.644     &  0.0387 &  0.6           & 1680   \\
\noalign{\smallskip}
{\it K}-cont     & 2.0975       &  0.0275 &  0.6       &  156    \\
\noalign{\smallskip}
H$_{2}$ 1-0 S(1) & 2.1219       &  0.0261 &  0.4        &  191    \\
\noalign{\smallskip}
H$_{2}$ 2-1 S(1) & 2.2465       &  0.0301 &  0.6       &  77    \\
\noalign{\smallskip}
CO 2-0 (bh) & 2.289        &  0.0279 &  0.5         &  126    \\
\hline
\end{tabular}
\end{table}

\subsection{Molecular data}

The molecular analysis was done at two different spatial scales: we used ALMA data of the complex molecular species CH$_{3}$OCHO and CH$_{3}$CN at high angular resolution and, to study the molecular environment of YSO-G29 at moderate angular resolutions we used \2 and \8 J=3--2 data obtained from a public database, and dedicated observations of HCN, HNC, HCO$^{+}$ J=4--3, and C$_{2}$H N=4--3 from ASTE.

\subsubsection{ALMA data}

Data cubes of the emission of methyl formate (CH$_{3}$OCHO) and methyl cianide (CH$_{3}$CN) with central frequencies at 226.71 and 239.01 GHz, respectively were obtained from the ALMA Science Archive\footnote{http://almascience.eso.org/aq/} (Project code: 2015.1.01312.S). The beam size of both data cubes is 0\farcs78 $\times$ 0\farcs60, which provides
a spatial resolution of about 0.02 pc ($\sim$4000 au) at the distance of 6.2 kpc. The velocity spectral resolution is 1.5 \ks. The Common Astronomy Software Applications (CASA) was used to handle these data. The task {\it imcontsub} was used to subtract the continuum from the spectral lines.

\subsubsection{\2 and \8 data}

The \2 and \8 J=3--2 data were obtained from the CO High-Resolution Survey (COHRS) and \3/\8 (J=3--2) Heterodyne Inner Milky Way Plane Survey (CHIMPS), two public databases of observations carried out with the 15-m James Clerck Maxwell Telescope (JCMT) in Hawaii.
The angular and spectral resolutions are 14\s~and 1 \ks~in the case of \2 \citep{dempsey13}, and 15\s~and 0.5 \ks~for the
\8 \citep{rigby16} providing a spatial resolution of about 0.4 pc at the distance of 6.2 kpc. The intensities of both data sets are on the $T_{A}^{*}$ scale, and the main beam efficiency $\eta_{\rm mb} = 0.61$ for the \2, and $\eta_{\rm mb} = 0.72$ for the \8 was used to convert $T_{A}^{*}$ to main beam brightness temperature, T$_{\rm mb} = T_{A}^{*}/\eta_{\rm mb}$  \citep{buckle09}.

\subsubsection{ASTE observations}

The observations of HCN, HNC, HCO$^{+}$ J=4--3, and C$_{2}$H N=4--3 were carried out on August 2019 with the 10-m ASTE telescope at the central frequencies 354.505, 362.630, 356.734 and 349.337 GHz, respectively. 
The DASH 345 GHz band
receiver was used with the FX-type spectrometer WHSF (bandwidth of 2018 MHz and spectral resolution of 1 MHz). The velocity resolution
is 0.86 \ks, and the beam size (FWHM) is 22\s, which provides a spatial resolution of about 0.6 pc at the distance of 6.2 kpc. 
The typical system temperature was about 300 K and the main beam efficiency $\eta_{mb} \sim 0.65$.
The observed pointing of the four molecular species was: $+$18\hh45\mm59.5\ss, $-$02\d45\m04.1\s~(J2000), with a pointing accuracy of about 3\s. The integration times were: 30, 36, 20, and 19 minutes for the HCN, HNC, HCO$^{+}$, and C$_{2}$H, respectively.

The data were reduced with NEWSTAR\footnote{Reduction software based on AIPS developed at NRAO, extended to treat single dish data
with a graphical user interface (GUI).}. The line base fitting was done using just a first order polynomial and the resulting rms noise level
was about 200 mK for all lines.

\section{Results}
\label{results}

Taking into account that the spatial scales corresponding to cores and clumps in which massive stars are born are $<0.2$ and $\sim0.5$ pc, respectively \citep{motte18}, our results are presented separated into both: in Sect.\,\ref{core} the results obtained from the Gemini near-IR observations and ALMA data at core scales and in Sect.\,\ref{clump}, the results obtained from the ASTE observations and the CO surveys at clumps scale.

\subsection{At core scales}
\label{core}

\begin{figure}[h]
\centering
\includegraphics[width=8.5cm]{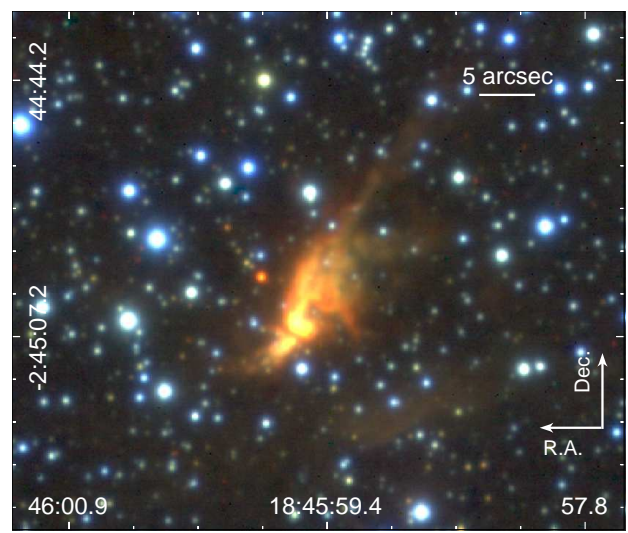}
\caption{Three-colour image of a 55\s$\times$45\s~region towards YSO-G29 with the {\it JHKs} broad-bands emission presented in blue, green, and red, respectively .}
\label{figjhk}
\end{figure}

\begin{figure}[h]
\centering
\includegraphics[width=7.8cm]{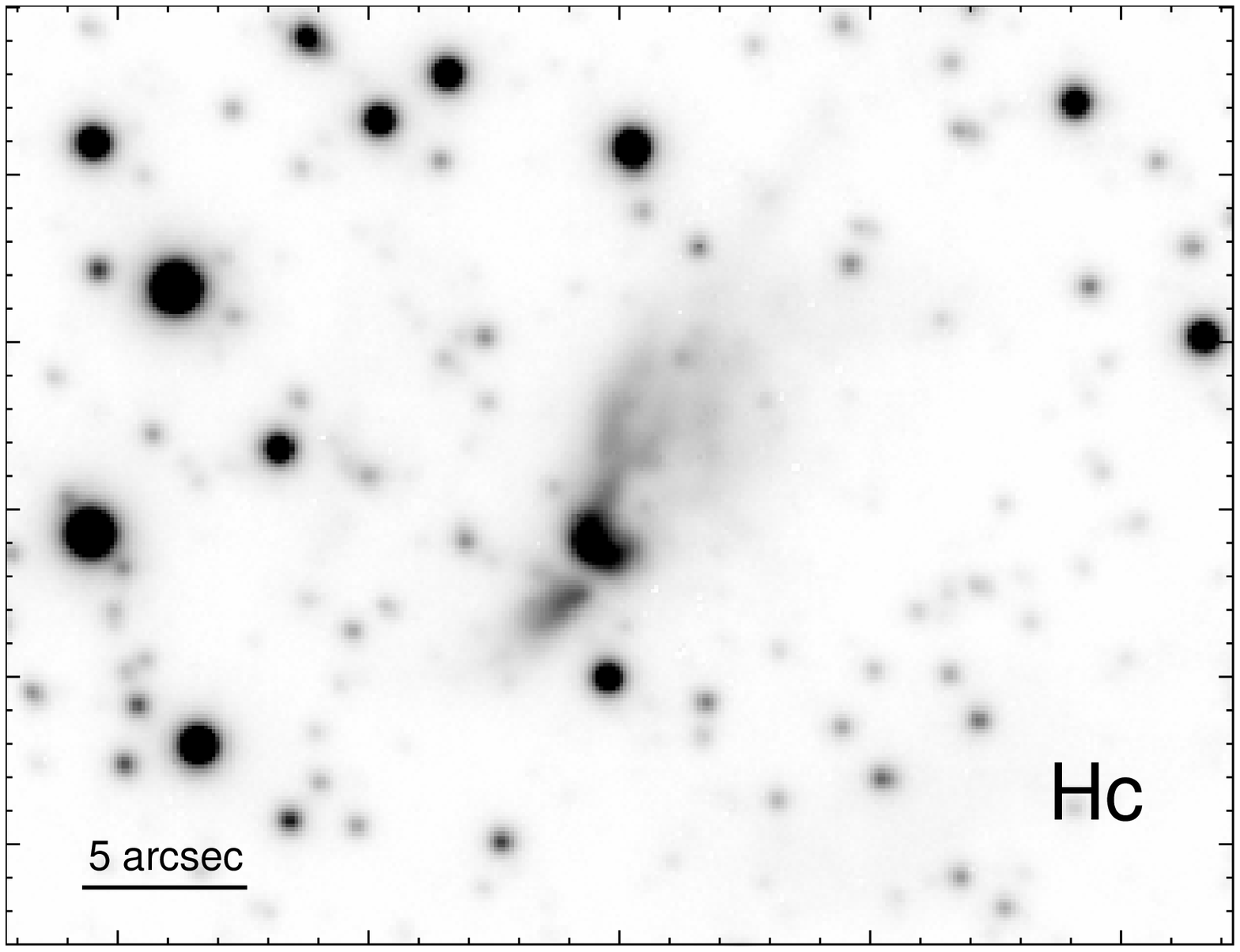}
\includegraphics[width=7.8cm]{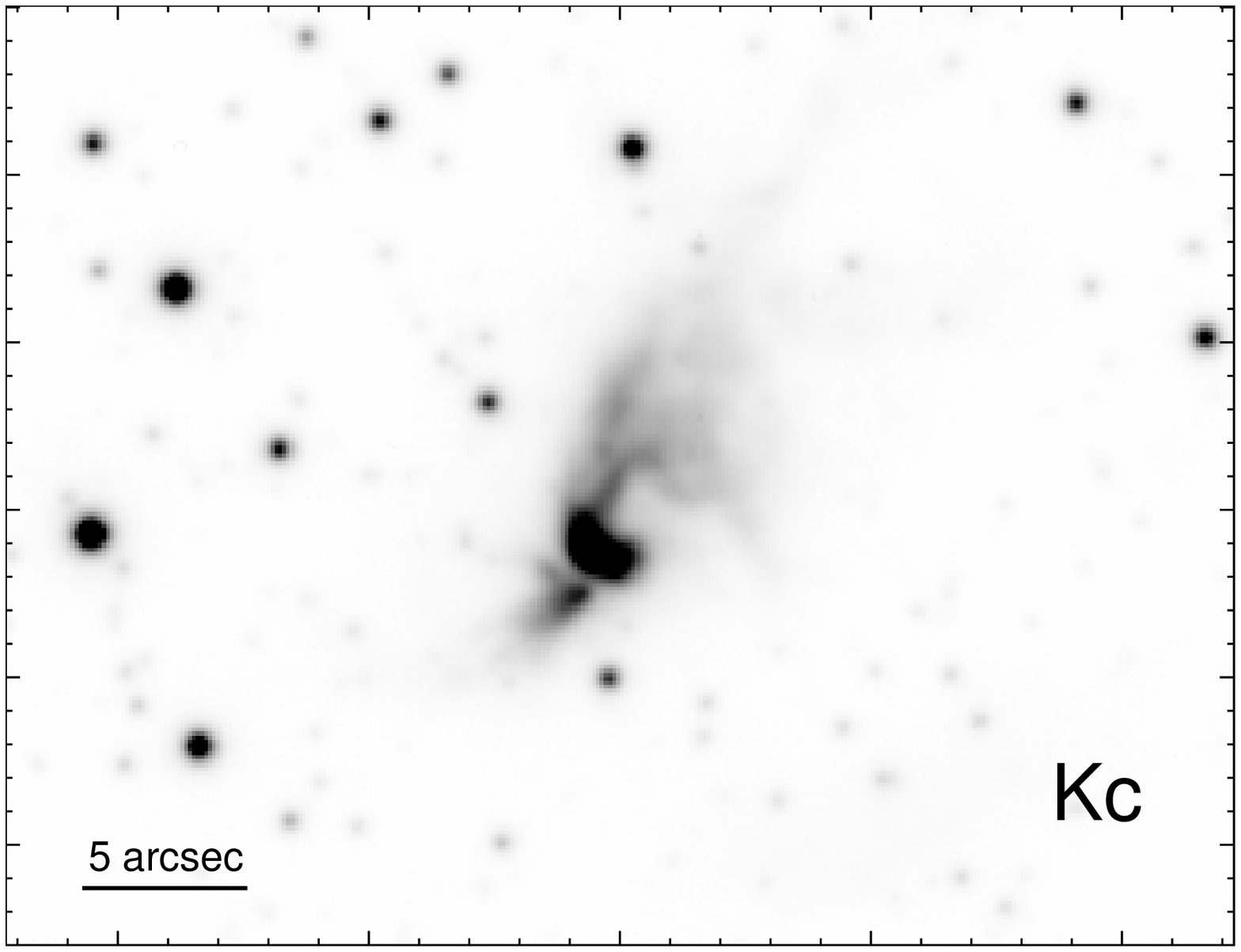}
\caption{Up: emission of the {\it H}-cont narrow-band (at 1.570 $\mu$m). Bottom: emission of the {\it K}-cont narrow-band (at 2.097 $\mu$m).}
\label{kchc}
\end{figure}

\begin{figure*}[h!]
\centering
\includegraphics[width=7cm]{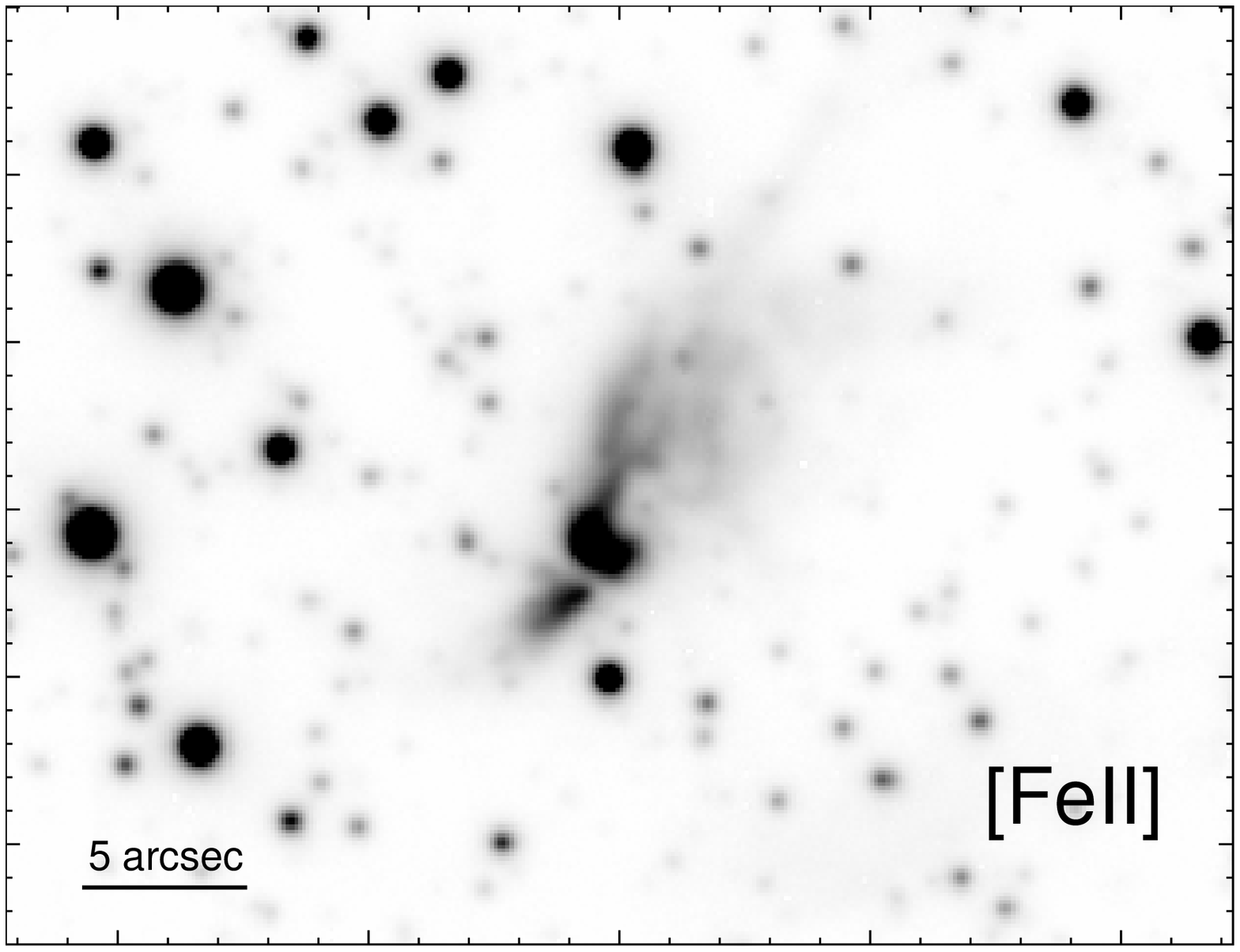}
\includegraphics[width=7cm]{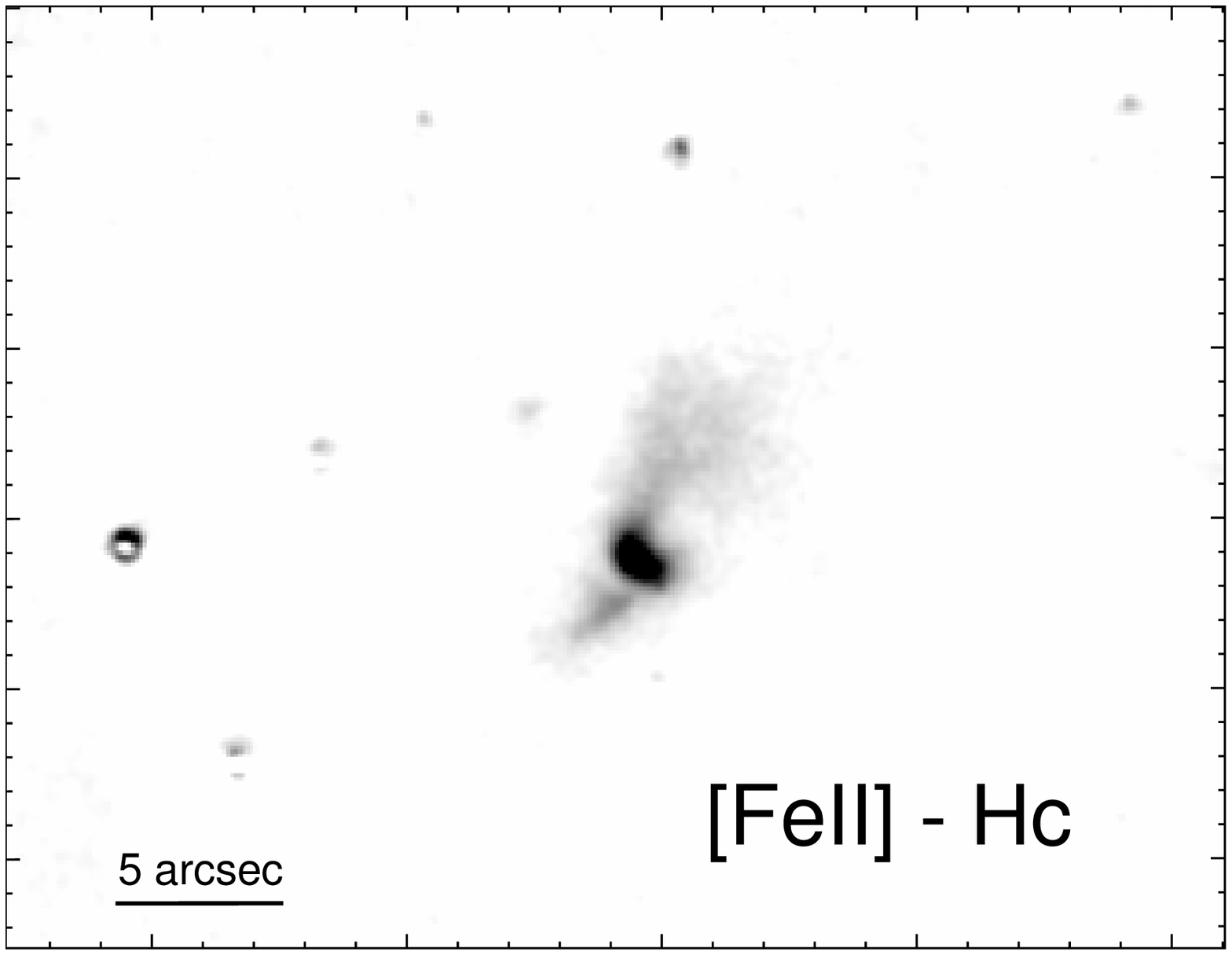}
\includegraphics[width=7cm]{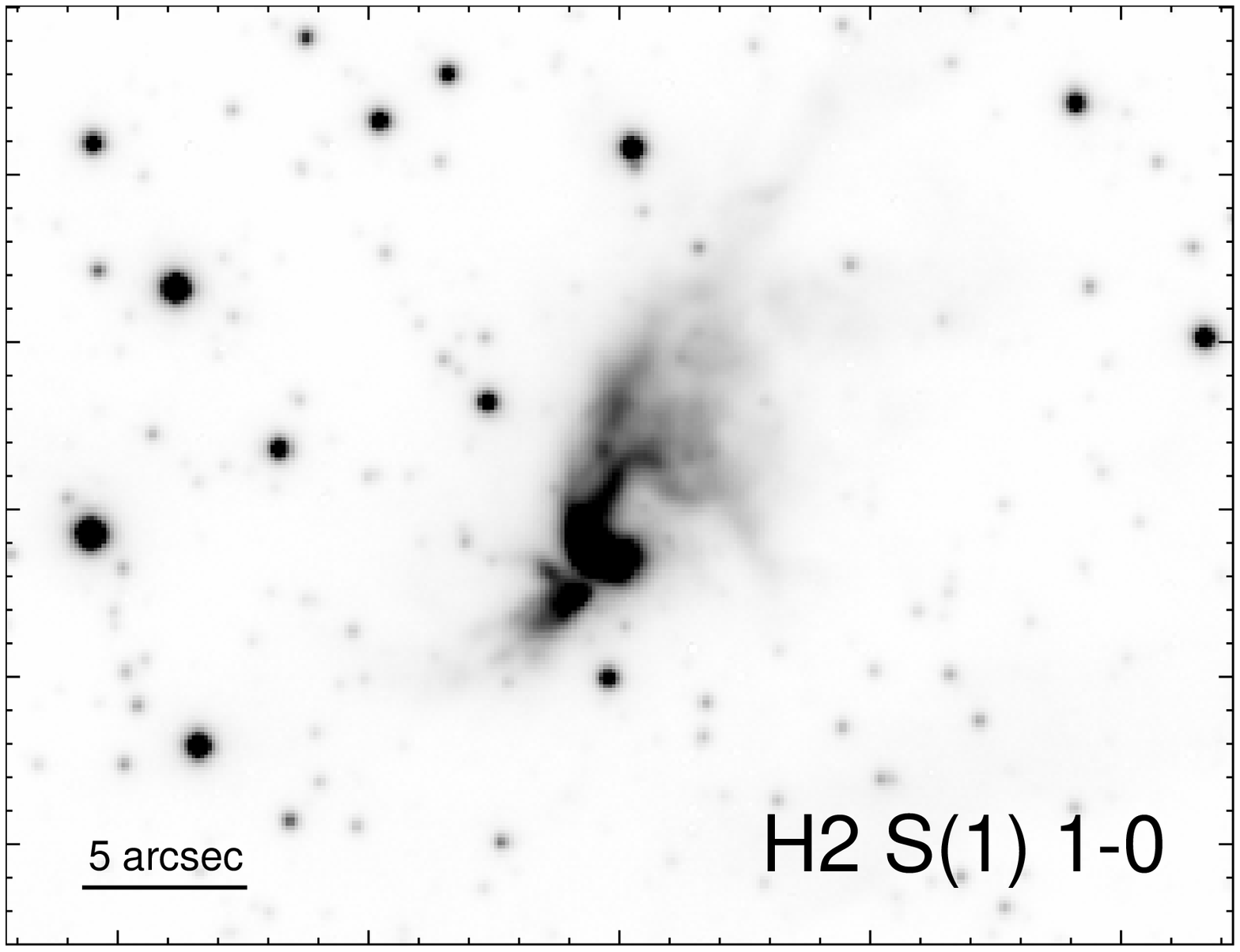}
\includegraphics[width=7cm]{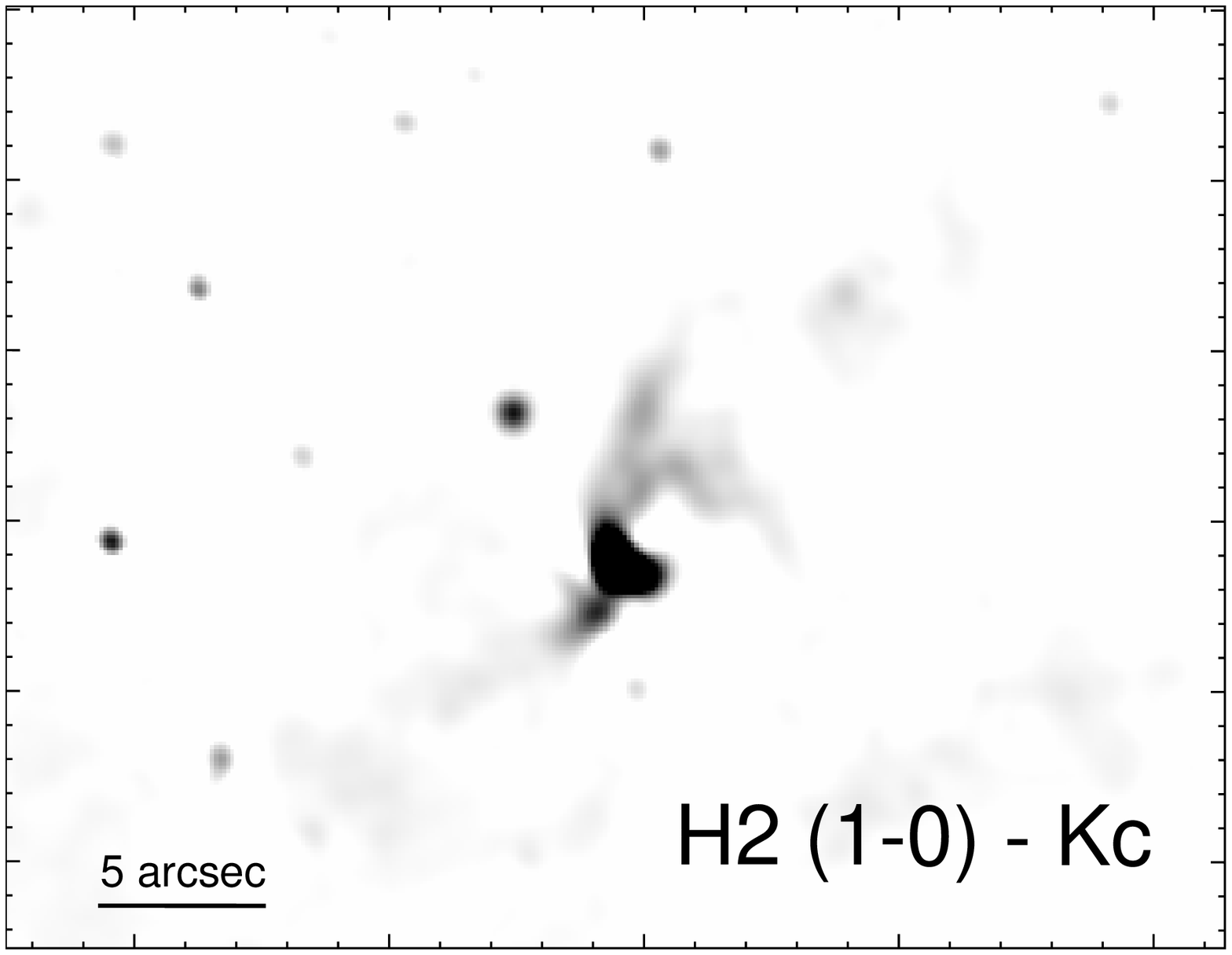}
\includegraphics[width=7cm]{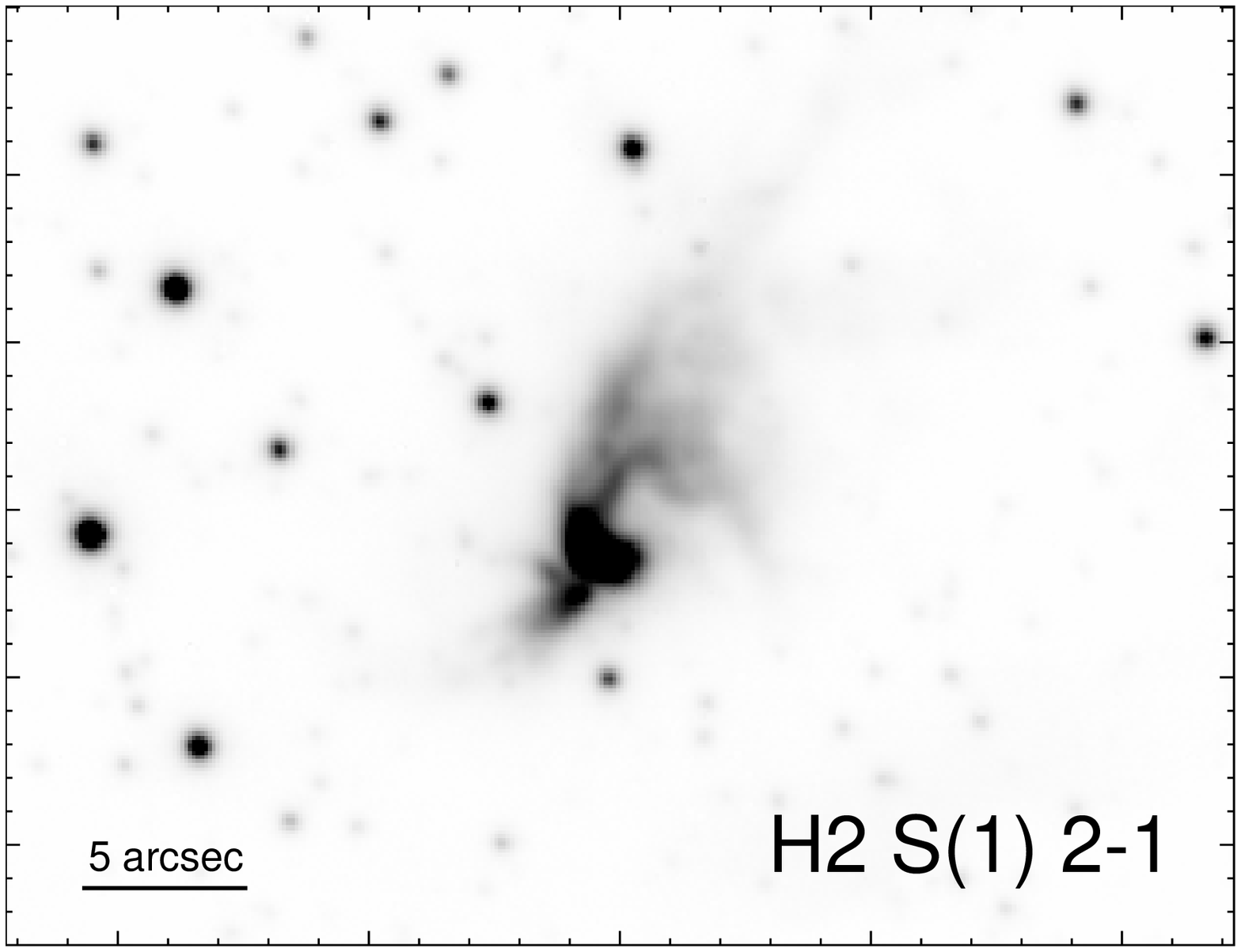}
\includegraphics[width=7cm]{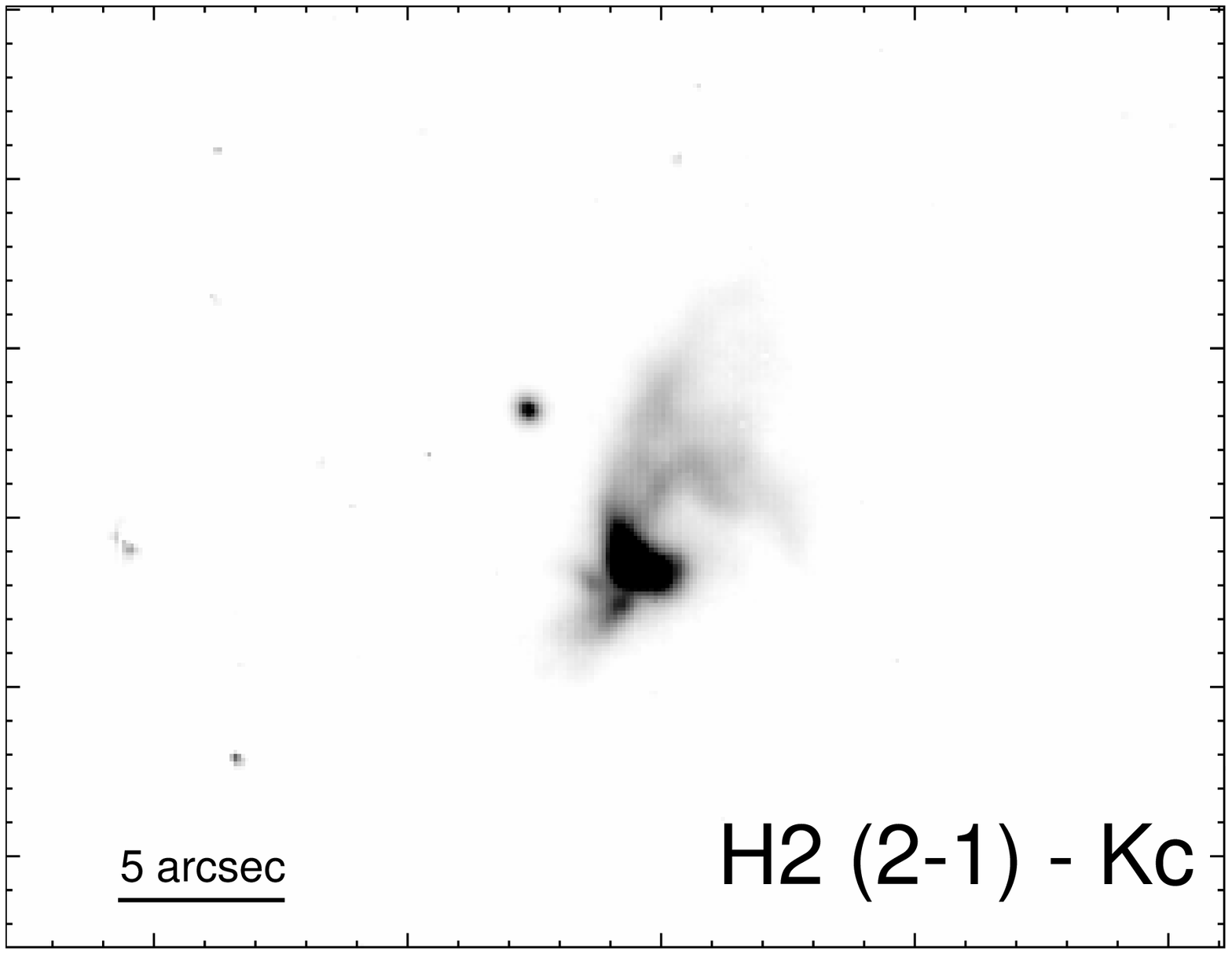}
\includegraphics[width=7cm]{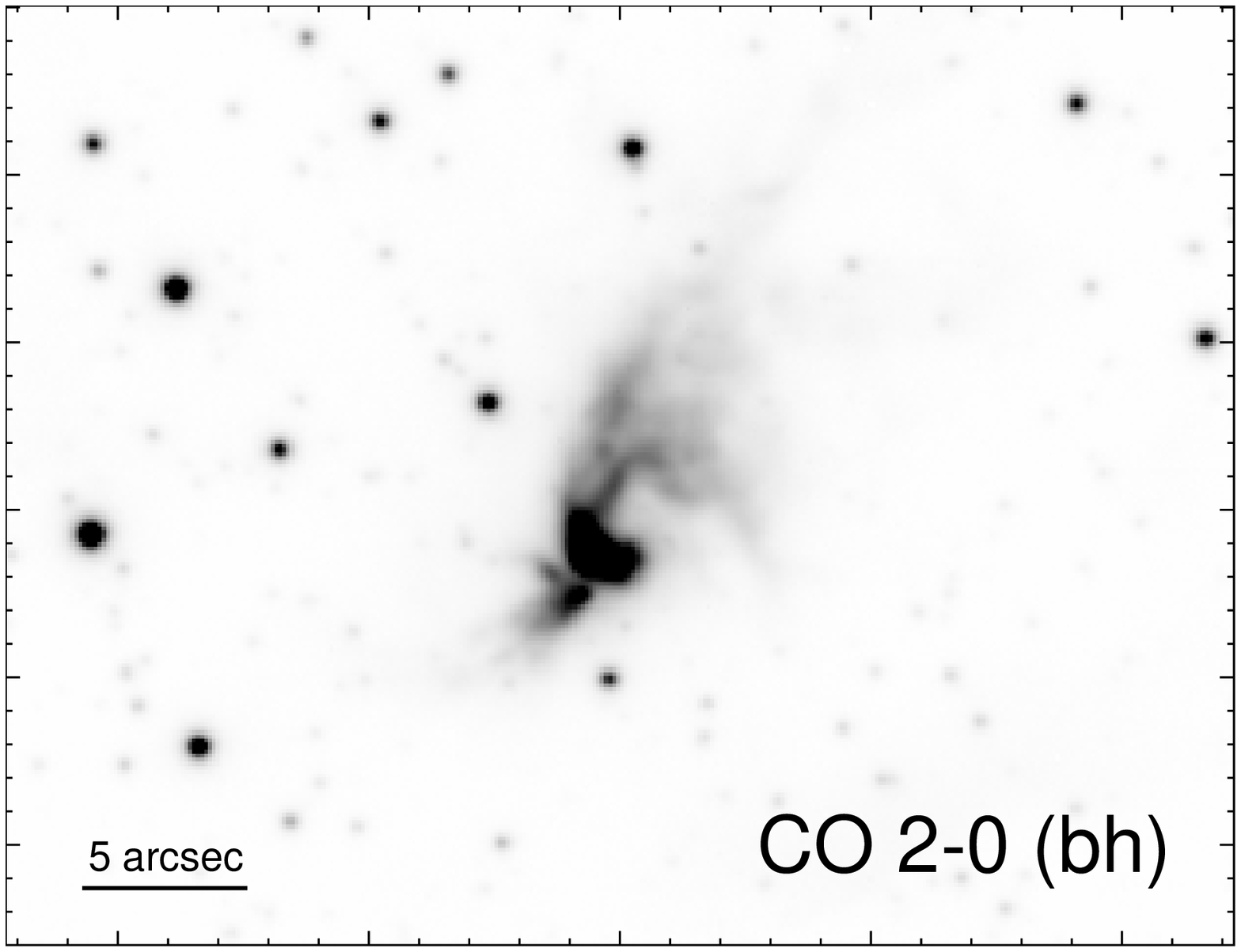}
\includegraphics[width=7cm]{white.ps}
\caption{From top to bottom: emission of the [FeII] (at 1.644 $\mu$m), H$_{2}$ 1-0 S(1) (at 2.2139 $\mu$m), H$_{2}$ 2-1 S(1) (at 2.2465 $\mu$m), and CO 2-0 (bh) (at 2.289 $\mu$m) with continuum (left column) and continuum subtracted (right column) except for the CO 2-0 (bh).}
\centering
\label{lines}
\end{figure*}

Figure\,\ref{figjhk} displays in a three-colour image the emission in the {\it JHKs} broad-bands of a region of about 55\s$\times$45\s~towards YSO-G29. 
Regarding the narrow bands, Fig.\,\ref{kchc} shows the emission of the {\it H}-cont and {\it K}-cont bands, while Fig.\,\ref{lines}
displays the emission of the [FeII], H$_{2}$ 1-0 S(1) and 2-1 S(1), and CO 2-0 (bh) lines with continuum (left panel), and
continuum subtracted (right panel), except for the CO 2-0 (bh) as the resulting continuum subtracted image quality is poor and hence not reliable. 
In the continuum subtracted images some  residuals of point sources remained due to differences in the point spread functions (PSF) of the targets in the final stacked images. Such PSF differences may arise as the combined effect of field distortions in NIRI and the relatively large offsets ($\sim$ 42 arcsec) between images resulting from the dithering pattern applied when observing. These features mainly affects point sources but their effect at the scales of extended sources is diluted.

\begin{figure}[h]
\centering
\includegraphics[width=8cm]{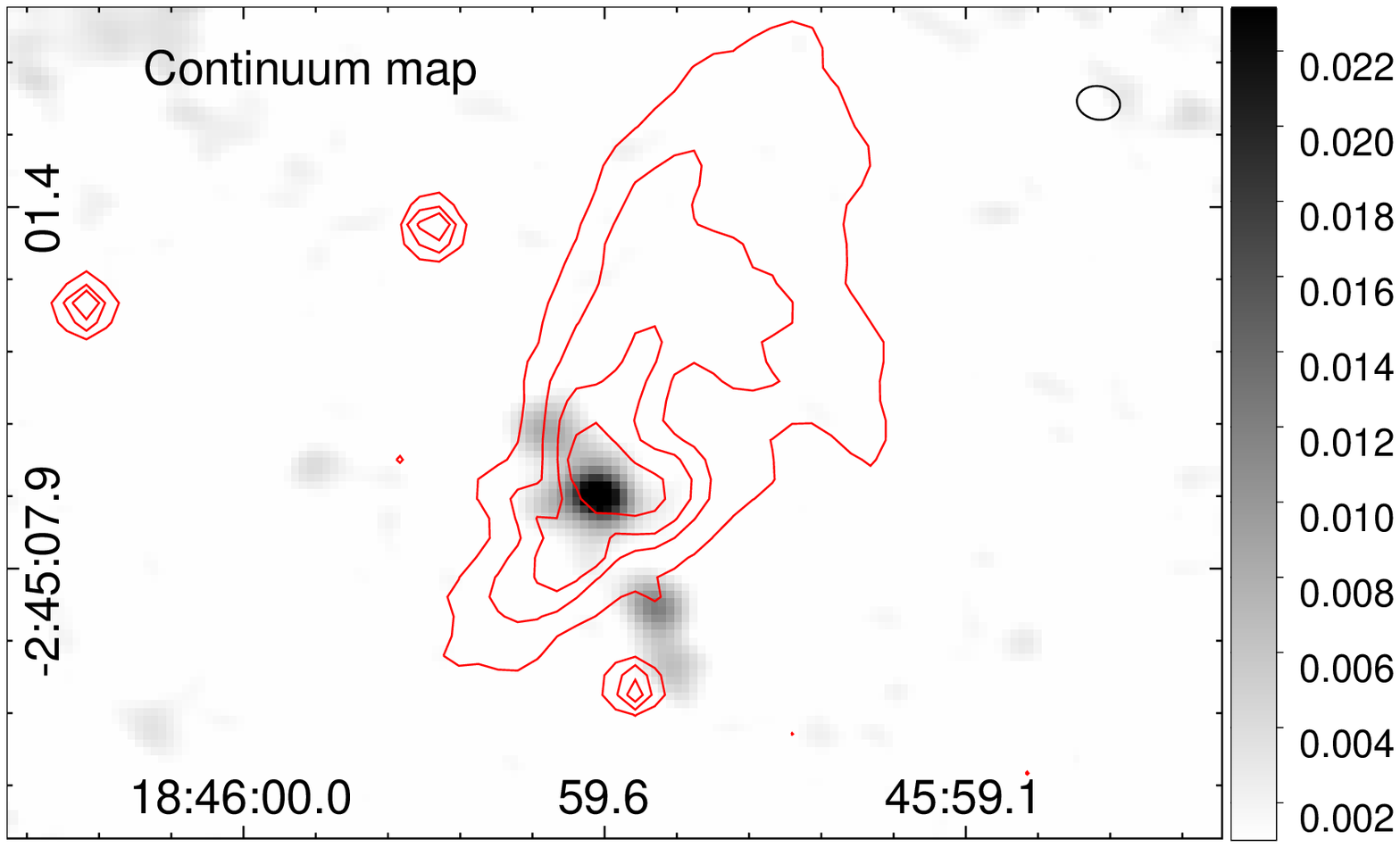}
\includegraphics[width=8cm]{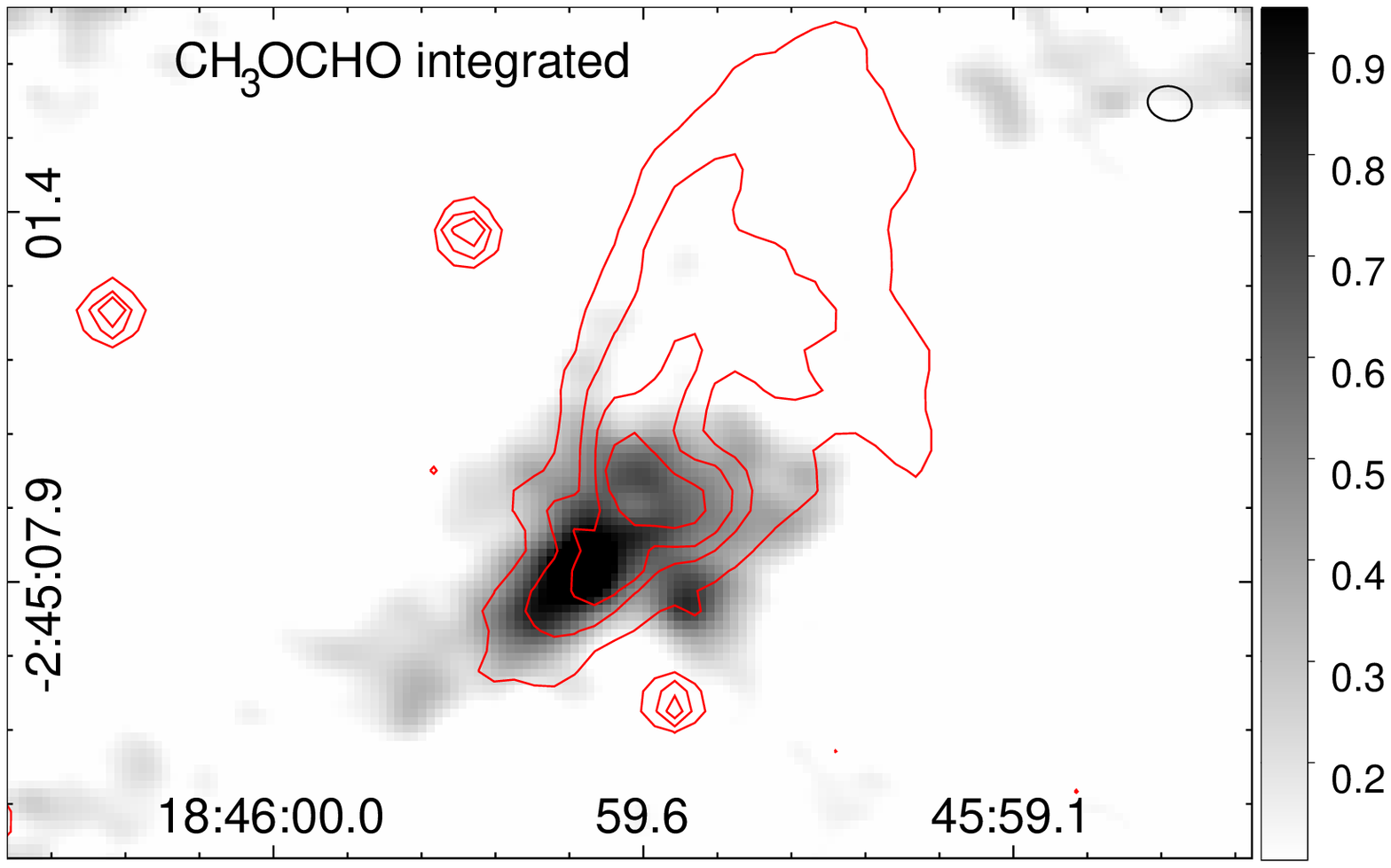}
\includegraphics[width=8cm]{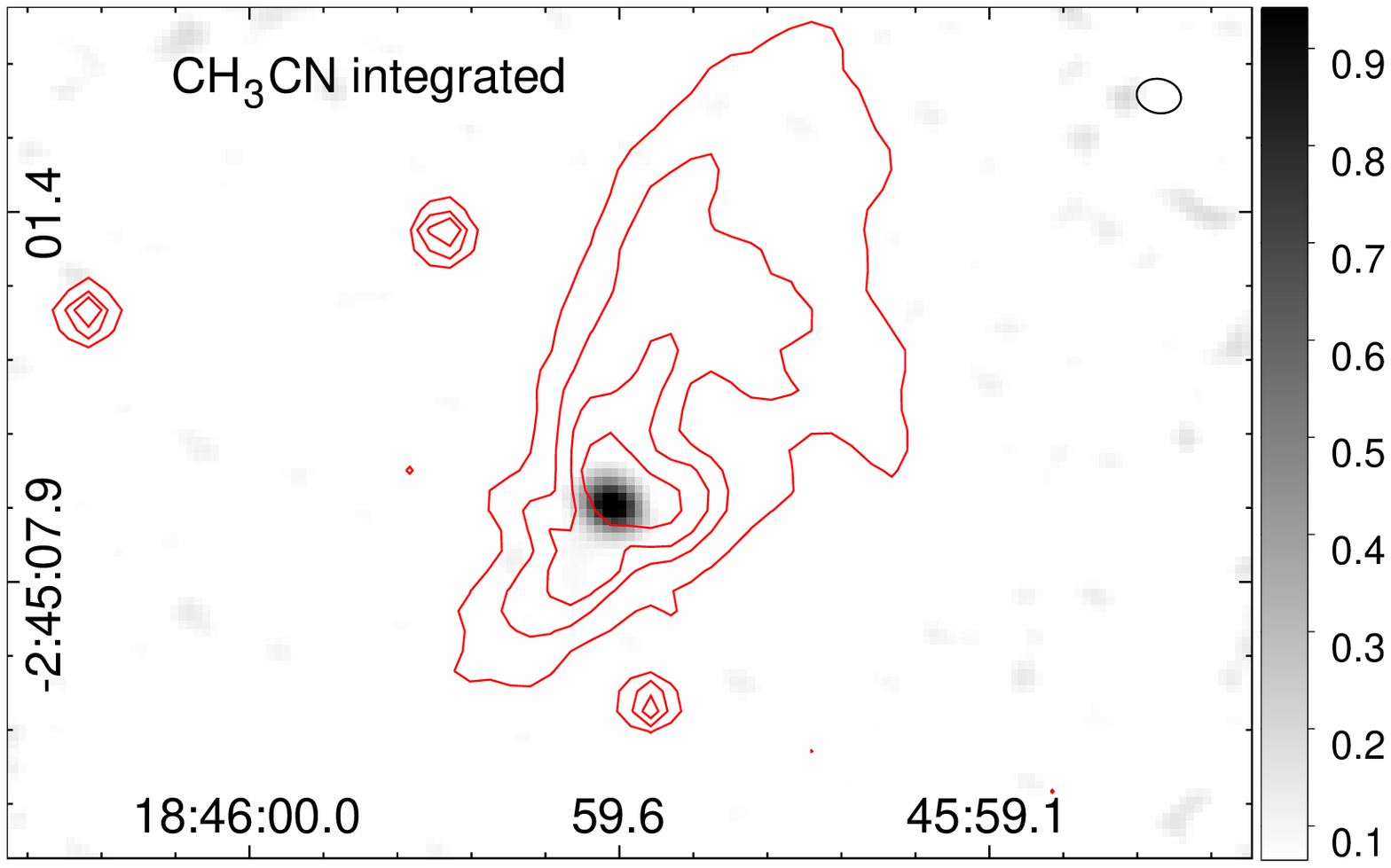}
\caption{ALMA continuum map at 1.3 mm of G29-YSO. The color scale is in Jy beam$^{-1}$, and the rms noise levels is 0.001 Jy beam$^{-1}$.
Middle and bottom panels: maps of the CH$_3$OCHO 20(2,19)--19(2,18)E emission integrated between 95 and 107 \ks, and the CH$_{3}$CN J=13--12 K=0 emission integrated between 98 and 105 \ks, respectively. The color scale is in Jy beam$^{-1}$ \ks. The rms noise levels are 0.10, and 0.05 Jy beam$^{-1}$ \ks, respectively. The beam of the ALMA data is included at the top right corner of each panel. Red contours correspond to the {\it Ks}-emission obtained with Gemini and are included for reference.
}
\label{almafig}
\end{figure}

Figure\,\ref{almafig} displays three maps obtained from the ALMA data: the continuum map at 1.3 mm and integrated emission maps of the complex molecules CH$_3$OCHO 20(2,19)--19(2,18)E and CH$_{3}$CN J=13--12 K=0. For comparison, contours of the near-IR {\it Ks}-emission obtained with Gemini are superimposed to each map. Table\,\ref{contparams}~presents the parameters obtained from the main core observed at 1.3 mm, i.e. the more intense core located almost at the center of the near-IR emission. The RA-dec. peak position is included in Cols. 1 and 2, the peak intensity and the integrated flux density corrected for primary beam response are presented in Cols. 3 and 4, the source size in arcsec deconvolved from beam is presented in Col. 5, and in Col. 6, the size in astronomical units considering a distance to YSO-G29 of 6.2 kpc.

\begin{table}
\tiny
\caption{Parameters of the central core observed from the ALMA continuum emission at 1.3 mm.}
\label{contparams}
\centering
\begin{tabular}{cccccc}
\hline\hline
RA(J2000) &   Dec.(J2000) & $I_{peak}$       & $S$     & $\theta_{s}$ &  $D_{s}$    \\
           &                & (mJy/beam) & (mJy)  &   (\s)      &    au     \\
\hline                 
18:45:59.566 & -2:45:5.518  &  26.2           & 51.6     &    0.7    &  4300 \\   
\hline
\end{tabular}
\end{table}

\begin{figure}[h]
\centering
\includegraphics[width=9cm]{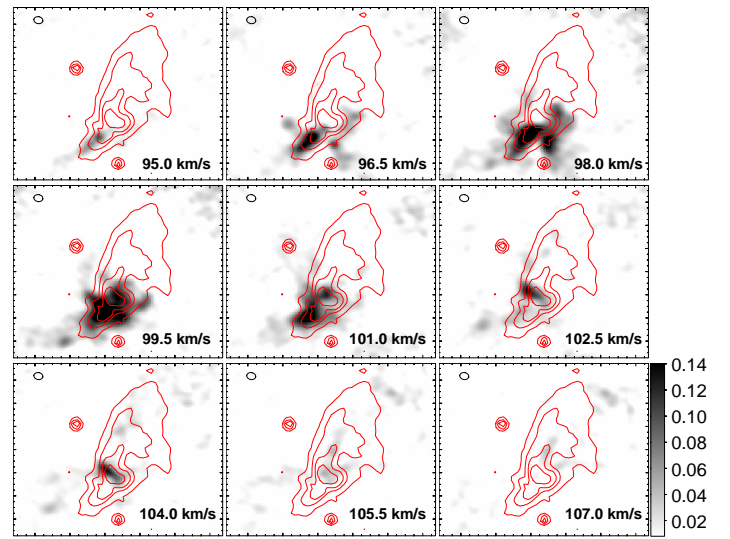}
\caption{Channel maps of the CH$_3$OCHO 20(2,19)--19(2,18)E. Red contours correspond to the {\it Ks}-emission obtained with Gemini and are included for reference. The color scale is in Jy beam$^{-1}$ and is shown at the last panel. The beam of the ALMA data is included at the top left corner in each panel.}
\label{chann1}
\end{figure}

\begin{figure}[h]
\centering
\includegraphics[width=9cm]{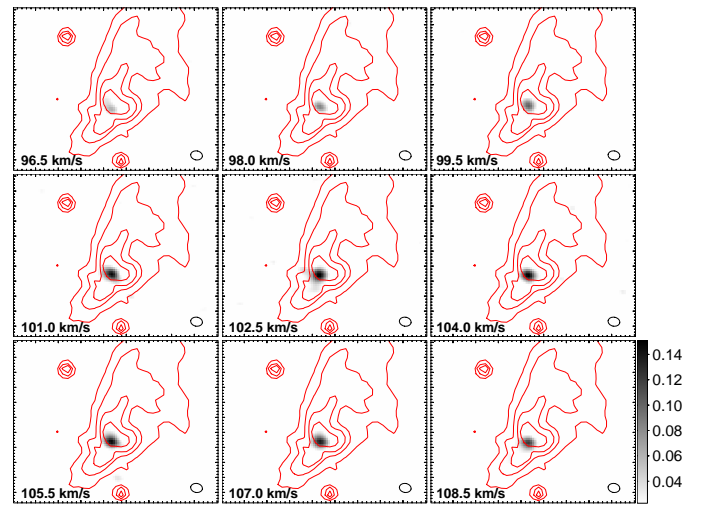}
\caption{Channel maps of the CH$_3$CN J=13--12 K=0. Red contours correspond to the {\it Ks}-emission obtained with Gemini and are included for reference. The color scale is in Jy beam$^{-1}$ and is shown at the last panel. The beam of the ALMA data is included at the bottom right corner in each panel. The emission, at panels 107.0 and 108.5 \ks, is very likely contaminated with the emission of the line with projection K=1.}
\label{chann2}
\end{figure}

To analyze the morphology of the molecular core along the line of sight and its kinematical features, Fig.\,\ref{chann1} presents a channel map of the CH$_3$OCHO 20(2,19)--19(2,18)E from 95 to 107 \ks~each 1.5 \ks, and the same is presented in Fig.\,\ref{chann2} for the CH$_{3}$CN J=13--12 K=0 line from 96.5 to 108.5 \ks, in which the last two panels may correspond to emission from the J=13--12 K=1 line. It is important to remark that the others CH$_{3}$CN K projections exhibit exactly the same morphology as K=0 in the integrated channel maps.

\begin{figure}[h]
\centering
\includegraphics[width=9cm]{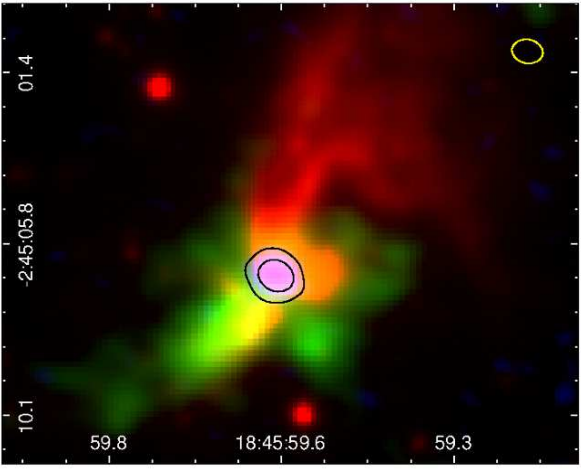}
\caption{Three-colour image composed of Gemini NIRI {\it Ks} emission (red) and the integrated emission of CH$_{3}$OCHO (green) and CH$_{3}$CN (blue) as presented in Fig.\,\ref{almafig}. The black contours are the integrated CH$_{3}$CN emission at levels 1.5 and 5.0 Jy beam$^{-1}$ \ks. The beam of the ALMA data is displayed at the top right corner.}
\label{comb}
\end{figure}

Figure\,\ref{comb} shows in a three-colour image the NIRI {\it Ks} emission in red and the integrated emission of the CH$_{3}$OCHO and CH$_{3}$CN in green and blue, respectively. Two contours of the integrated CH$_{3}$CN at 1.5 and 5.0 Jy beam$^{-1}$ \ks~levels are also included.

\subsubsection{Physical parameters from the CH$_{3}$CN emission}

\begin{figure}[h!]
\centering
\includegraphics[width=8.5cm]{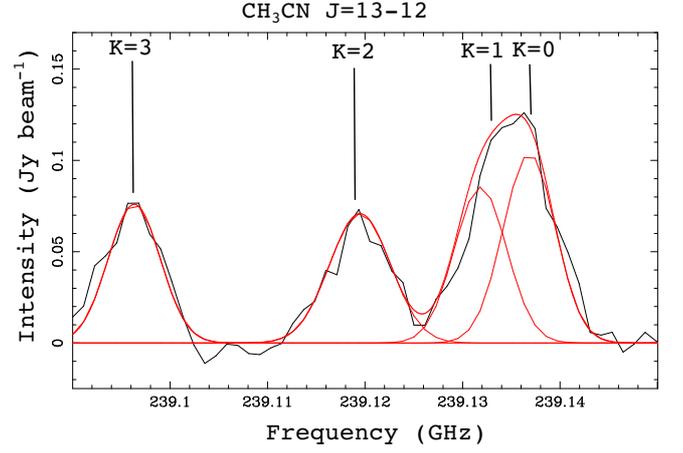}
\caption{CH$_{3}$CN J=13--12 averaged spectrum towards G29-YSO at rest frequency. The K projections of the J line are marked. The single or multiple-component Gaussian fits are shown in red.}
\label{ch3cnspect}
\end{figure}

\begin{figure}[h!]
\centering
\includegraphics[width=8.5cm]{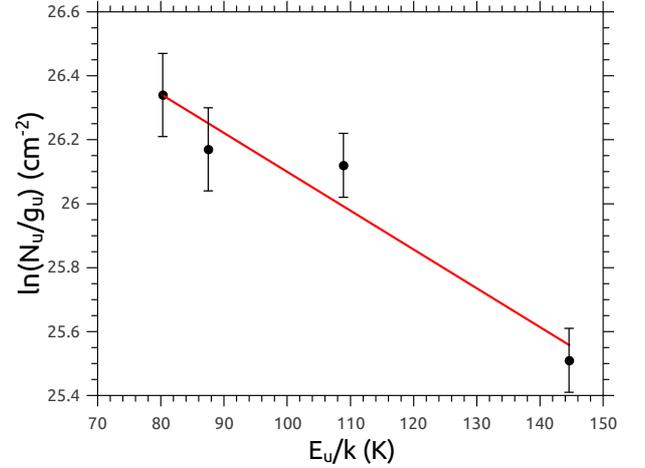}
\caption{Rotation diagram of the CH$_{3}$CN J=13--12 obtained from K=0, 1, 2, and 3 projections. The red line is the linear fitting.}
\label{trot}
\end{figure}

Figure\,\ref{ch3cnspect} shows an averaged spectrum of the CH$_{3}$CN J=13--12 towards G29-YSO. The K projections of the transition are marked. The relevant parameters from the Gaussian fittings presented in Table\,\ref{ch3cnKs} are the central frequency, the peak intensity ($I_{peak}$), and the integrated flux density ($W$). The energy of the upper level ($E_{u}$) and the line strength of the transition multiplied by the dipolar moment ($S_{ul}\mu^{2}$) are also included. The integrated flux densities were used to construct the rotational diagram (RD) presented in Fig.\,\ref{trot}. 
By assuming LTE conditions, optically thin lines, and a beam filling factor equal to the unity, we can estimate the rotational excitation temperature ($T_{rot}$) and the column density of the CH$_3$CN from the rotation diagram (RD) analysis \citep{turner91}. This analysis is based on a derivation of the Boltzmann equation,

\begin{equation}
{\rm ln}\left(\frac{N_u}{g_u}\right)={\rm ln}\left(\frac{N_{tot}}{Q_{rot}}\right)-\frac{E_u}{kT_{rot}},   \label{RotDia} 
\end{equation}

\noindent where $N_u$ represents the molecular column density of the upper level of the transition, $g_u$  the total degeneracy of the upper level, $N_{tot}$ the total column density of the molecule, $Q_{rot}$ the rotational partition function, and $k$ the Boltzmann constant. 

Following \citet{miao95}, for interferometric observations, the left-hand side of Eq.\,\ref{RotDia} can also be estimated by,

\begin{equation}
{\rm ln}\left(\frac{N_u^{obs}}{g_u}\right)={\rm ln}\left(\frac{2.04 \times 10^{20}}{\theta_a \theta_b}\frac{W}{g_kg_l{\nu_0}^3S_{ul}{\mu_0}^2}\right)-\frac{E_u}{kT_{rot}},   
\label{RD} 
\end{equation}

\noindent where $N_u^{obs}$ (in cm$^{-2}$) is the observed column density of the molecule under the conditions mentioned before, $\theta_a$ and $\theta_b$ (in arcsec) are the major and minor axes of the clean beam, respectively, $W$ (in Jy beam$^{-1}$ \ks) is the integrated intensity of each K, $g_k$ is the K-ladder degeneracy, $g_l$ is the degeneracy due to the nuclear spin, $\nu_0$ (in GHz) is the rest frequency of the transition, $S_{ul}$ is the line strength of the transition, and $\mu_0$ (in Debye) is the permanent dipole moment of the molecule.  The free parameters, ($N_{tot}/Q_{rot}$) and $T_{rot}$ were determined by a linear fitting of Eq.\,\ref{RotDia} (see Fig.\,\ref{trot}). We derive a $T_{rot}$ of about 82~K and, using a tabulated value for $Q_{rot}$ at this temperature, we obtain a CH$_3$CN column density of about $10^{15}$ cm$^{-2}$.

\begin{table}
\tiny
\caption{Tabulated and Gaussian fittings parameters (see Fig.\,\ref{ch3cnspect}) for the K projections of the CH$_{3}$CN J=13--12 transition. The center frequency takes into consideration the collapsed hyperfine components for each projection (spacing smaller than channel width).}
\centering
\begin{tabular}{c c c c c c}
\hline\hline
    K     &   Frequency &  $E_u$ & $S_{ul}\mu^2$ & $I_{peak}$   &   $W$   \\
          &   (GHz)     & (K)      &   (Debye$^2$)  &   (Jy/beam)  &  (Jy/beam \ks)    \\
\hline                 
    0     &  239.137   &  80.3 & 199.1 & 0.10   &  0.75  \\
    1     &  239.133   &  87.5 & 197.9 & 0.08   &  0.63  \\
    2     &  239.119   & 108.9 & 194.3 & 0.07   &  0.59  \\
    3     &  239.096   & 144.6 & 188.5 & 0.07   &  0.62  \\
 \hline
\end{tabular}
\label{ch3cnKs}
\end{table}

\subsubsection{Mass and density of the core}

We estimate the mass of gas of the central core from the dust continuum emission at 1.3~mm following  \citet{kau08},

\begin{eqnarray}
M_{gas}=0.12~{\rm M_\odot} \left[exp\left(\frac{1.439} {(\lambda/{\rm mm})(T_{dust}/10~{\rm K})}\right)-1\right] \\ \nonumber \times\left(\frac{\kappa_{\nu}}{0.01~{\rm cm}^2~{\rm g}^{-1}}\right)^{-1}\left(\frac{S_{\nu}}{\rm Jy}\right)\left(\frac{d}{100~{\rm pc}}\right)^2\left(\frac{\lambda}{\rm mm}\right)^3
\label{massdust}
\end{eqnarray}

\noindent where $T_{dust}$ is the dust temperature and $\kappa_{\nu}$ is the dust opacity per gram of matter at 1.3~mm for which we adopt the value of 0.01~cm$^2$g$^{-1}$ \citep{kau08, osse94}.
Assuming LTE conditions ($T_{kin}$=$T_{rot}$), and that the gas and dust are thermally coupled ($T_{dust}$=$T_{kin}$), and considering the integrated flux intensity $S_{\nu}$ = 0.051~Jy at 1.3~mm (see Table \ref{contparams}), we obtain $M_{gas} \sim 8$~M$_\odot$. Hence, assuming spherical geometry we can derive a volume density of n(H$_2$)$ \sim 1 \times 10^6$ cm$^{-3}$.

Additionally, neglecting contributions from magnetic field and surface pressure, we estimate the virial mass using the CH$_{3}$CN emission and the following equation:
\begin{equation}
M_{vir}/M_\odot = k~R/pc~(\Delta {\rm v}/{\rm km~s^{ -1}})^{2}
\label{vir}
\end{equation}
where $k=190$ by assuming a density distribution $\propto r^{-1}$ \citep{mac88}, $\Delta$v the line velocity width (FWHM) of the CH$_{3}$CN emission obtained from the gaussian fittings presented in Fig.\,\ref{ch3cnspect}, estimated to be about 6.7 \ks~(the $\Delta$v of all K projections of the J=13--12 transition goes from 5.5 to 8.0  \ks), and $R$ is the radius, about 0.01 pc (from the source size presented in Table\,\ref{contparams}). Thus, we obtain $M_{vir} \sim 70$ \msol.

\subsection{At clump scales}
\label{clump}

\begin{figure}[h!]
\centering
\includegraphics[width=8.5cm]{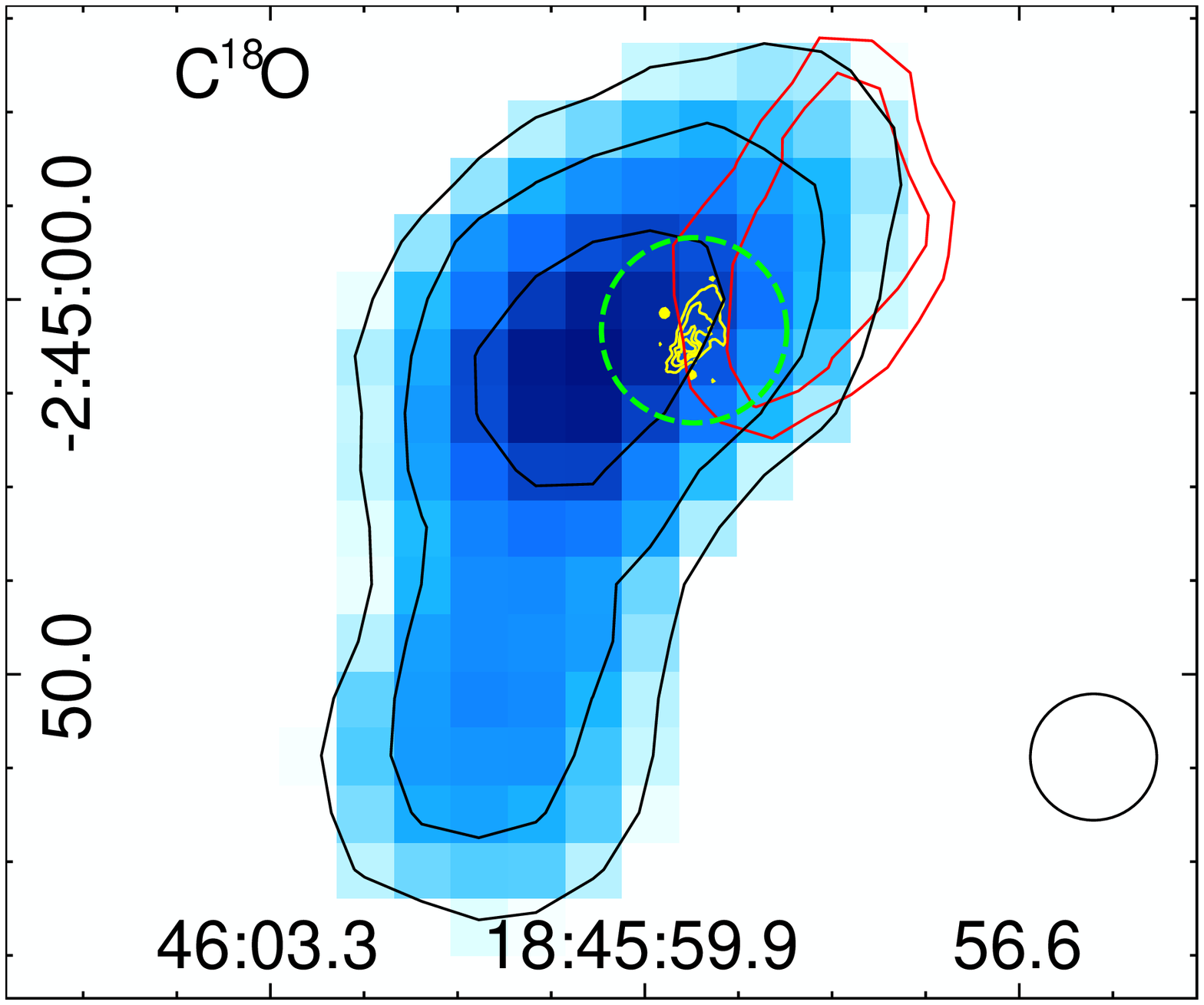}
\caption{\8 J=3--2 emission integrated between 99 and 104.5~\ks. The contour levels are 5, 7, and 10~K \ks. The red contour corresponds to the \2 J=3--2 emission integrated between 104 and 112~\ks. The red contour levels are 40 and 50~K \ks. The beam is included at the bottom right corner. The yellow contours correspond to the near-IR {\it Ks}-emission obtained with Gemini. The dashed circle represents the pointing and the beam of the ASTE observations.}
\label{figred}
\end{figure}

\begin{figure}[h]
\centering
\includegraphics[width=8cm]{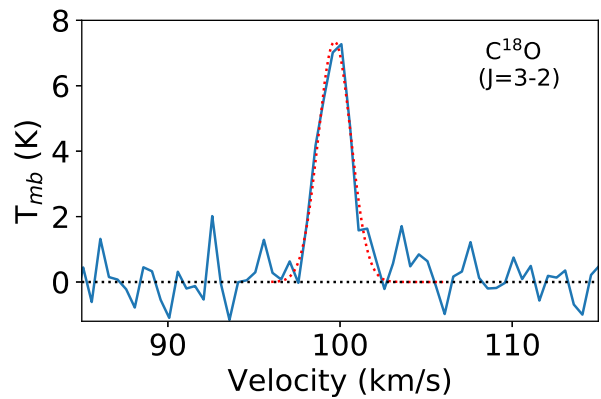}
\caption{\8 J=3--2 spectrum from the JCMT data at the G29-YSO position. The red dotted lines are the Gaussian fittings.}
\label{c18ospect}
\end{figure}

To study the cloud and the clump in which G29-YSO is embedded, in Fig.\,\ref{figred} we present the integrated \8 J=3--2 emission between 98.0 and 104.5~\ks~(the velocity range in which the emission of the \8 extends). Following previous works (see Sect.\,\ref{intro}) the \2 J=3--2 emission integrated between 104 and 112~\ks~(the red wing as observed in this line) is displayed in red contours, showing the already reported red molecular outflow. In addition, in Fig.\,\ref{c18ospect} we present the \8 J=3--2 spectrum obtained towards the G29-YSO position. By integrating the \2 in the velocity range corresponding to the blue wing suggested by \citet{yang18}, it is possible to distinguish a molecular feature associated with the southern portion of the elongated cloud observed in the \8 line, which is coincident with another cold dust clump as observed in ATLASGAL and radio continuum sources (NVSS J184603-024541 and NVSS J184601-024601), showing that it should be molecular gas related to another active region within the same complex. Thus, this blue component in the \2 cannot be considered as a molecular outflow.

\begin{figure}[h]
\centering
\includegraphics[width=7cm]{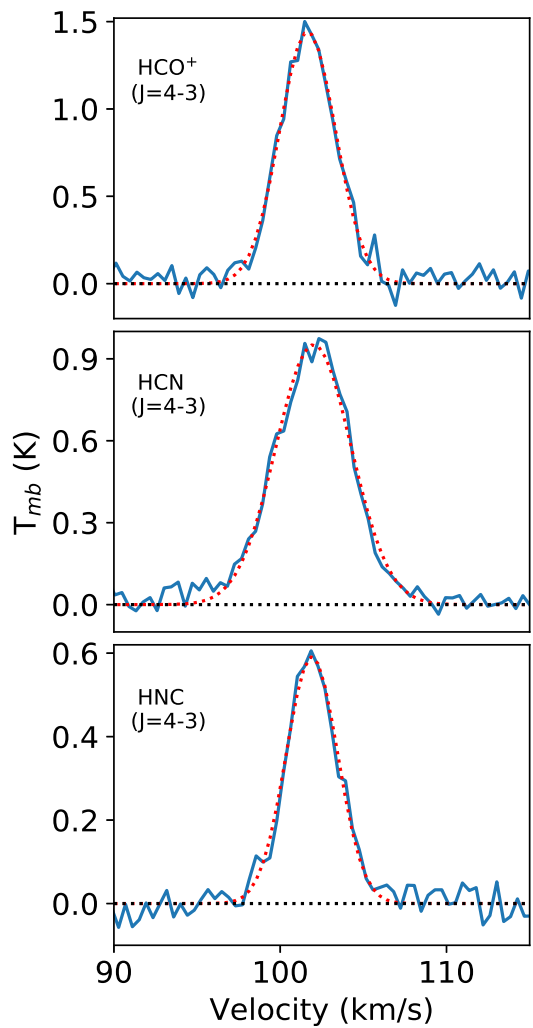}
\caption{HCO$^{+}$, HCN, and HNC J=4--3 observed with ASTE towards G29-YSO. The red dotted lines are the Gaussian fittings.}
\label{hcspect}
\end{figure}

\begin{figure}[h]
\centering
\includegraphics[width=8cm]{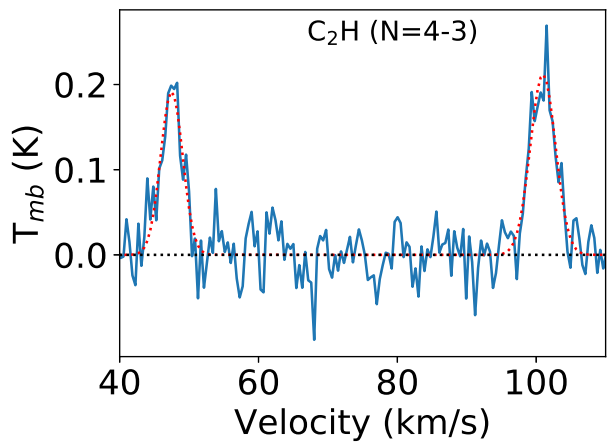}
\caption{C$_{2}$H spectrum observed with ASTE towards G29-YSO. The red dotted lines are the Gaussian fittings.}
\label{c2hspect}
\end{figure}

\begin{table}
\caption{Parameters derived from the  Gaussian fitting to the \8, C$_2$H, HCN, HNC, and HCO$^{+}$ spectra shown in Figs.\ref{c18ospect},\ref{hcspect} and \ref{c2hspect}.}
\label{molec}
\centering
\begin{tabular}{c c c c c}
\hline\hline
    Molecule        &   T$_{mb}$    &   V$_{LSR}$     & $\Delta$v     & $\int T_{mb} dv$    \\
                    &   [K]         &   [\ks]         & [\ks]         & [K \ks]     \\
\hline       
    C$^{18}$O       &   7.32        &   100.05        &   2.88        &  17.41    \\
    C$_2$H          &   0.18        &   --            &   4.06        &   0.76    \\
                    &   0.20        &   100.98        &   4.56        &   0.96    \\
    HCN             &   0.95        &   102.01        &   5.33        &   5.41    \\
    HNC             &   0.59        &   101.91        &   3.64        &   2.29    \\
    HCO$^{+}$       &   1.44        &   101.65        &   3.97        &   6.09    \\
    
\hline
\end{tabular}
\end{table}

Figure\,\ref{hcspect} shows the spectra from the emission of the HCO$^{+}$, HCN, and HNC. Fig.\,\ref{c2hspect} shows the spectrum of C$_{2}$H, as observed with ASTE towards the YSO position (dashed green circle in Fig.\,\ref{figred}). 
Table\,\ref{molec} lists the parameters obtained from the Gaussian fitting to each spectrum presented in Figs.\ref{c18ospect},\ref{hcspect} and \ref{c2hspect}. The C$_{2}$H presents two peaks due to its fine structure components. The central frequency to observe this molecular species was set at 349.337 GHz, which corresponds to the N=4-3 J=9/2-7/2 F=5-4 transition. While each C$_{2}$H peak was fitted by one gaussian, it is important to remark that one peak should correspond to the blended C$_{2}$H (N=4–3) J=9/2–7/2 F=5–4 and 4–3 lines (from which the systemic velocity of 100.98 \ks~is obtained), and the other to the blended (N=4–3) J=7/2–5/2 F=4–3 and 3–2 lines (see for instance the NIST data base\footnote{http://www.nist.gov/pml/data/micro/index.cfm}). The same occurs with the HCN J=4--3 line, which also is affected by the hyperfine splitting. The HCN/HNC integrated ratio is $\sim 2.3$.

\section{Discussion}

As done in Sect.\,\ref{results}, the discussion of the results are presented separated into both (core and clump) spatial scales analyzed.

\subsection{Core scales}

The near-IR bands show that YSO-G29 is composed by two nebulosities separated by a dark lane (see Figs.\,\ref{figjhk},\,\ref{kchc},  and\,\ref{lines}). 
The dark lane could be due to the presence of an accretion disk or a rotating toroid which may hide the protostar(s), as it is usually found towards YSOs observed mainly edge-on \citep{beltran16,furu08}. The near-IR emission is highly asymmetric, being the northern nebulosity extended and open, while the southern counterpart is smaller. Overall, the nebulosities seem to be composed by some jet-like structures, suggesting a
scenario consistent with the typical disk-jets systems observed along decades generally towards nearby 
YSOs (e.g. \citealt{pad99}) and in agreement with the description given for this source in the RMS survey of \citet{lumsden13} based on UKIDSS near-IR data. However, the high resolution of NIRI's data allows to distinguish some intriguing features. The jet-like structure extending towards the northwest is divided into two branches. One branch curves towards the west and slightly to the south, generating a M-shaped structure (see mainly the {\it Ks} emission displayed in red in Figs.\,\ref{figjhk} and \ref{comb}). The other branch extends mainly straight to the north. Diffuse emission is observed between both branches. On the other hand, the southern jet-like feature leans more sharply to the southeast. 

It is well known that the near-IR emission towards this kind of objects arises mainly from scattered light in the material surrounding the protostar(s), emission from warm dust plus atomic and molecular emission lines (e.g. \citealt{muze13,bik06,bik05,reip00}). The geometry of the near-IR emission presented here is quite similar to that detected towards the protostar L54361 \citep{muze13} and, as the authors proposed, it is probably produced by the scattered light in cavities carved out by one or more jets on an infalling envelope of material.

Most of the structures that are observed in the {\it Ks}-band are also present in the H$_{2}$ 1--0 and 2--1 S(1) as Fig.\,\ref{lines} (right panel) shows.  
The near-IR emission from H$_{2}$ can be generated by either thermal emission in shock fronts or fluorescence excitation by non-ionizing UV photons \citep{m-h08}. As those authors mention, in high density regions (as it is this case) a combination of 1-0 (S1)/2-1 (S1) and 1-0 S(1)/3-2 S(3) ratios is necessary to assess the H$_{2}$ excitation mechanism. Our near-IR data set is not complete for specifying this and even though in the NIRI's images there is not evidence of a H$_{2}$ well collimated structure as usually found towards jets (e.g. \citealt{dewan15}), considering that jets could be curved in certain conditions we cannot discard shocks contribution for the observed H$_{2}$ emission. Additionally, given that the H$_{2}$ radiative excitation, and hence its emission, should be due to UV photons impinging the molecular gas in cavities carved out by the 
YSO jets and/or winds \citep{frank14}, it makes sense to conclude that the observed features in the H$_{2}$ are indeed due to the activity of a protostar(s) likely hidden in the dark lane.
The [FeII] emission presents, in general, a similar morphology as the other bands, but the jet-like structures are much less conspicuous  and more diffuse. The origin of this emission can be due to knots and shock fronts along the regions where the jets propagate \citep{bally07,davis11}.

From the analysis of the ALMA data (see Figs.\,\ref{almafig},\,\ref{chann1},\,\ref{chann2}, and\,\ref{comb}) we find a well defined core mapped in the continuum emission at 1.3 mm lying almost at the position of the dark lane of the near-IR emission (slightly displaced to the north) and other weaker cores towards the northeast and southwest. The main core has a size of about 4300 au, which is in agreement, together with the measured flux and peak intensity (see Table\,\ref{contparams}) with the values obtained towards the hot core G31-NE \citep{beltran18}. The CH$_{3}$CN emission is in perfect spatial coincidence with the central core as observed in 1.3 mm continuum. The CH$_{3}$OCHO emission is more extended and presents some knots and filament-like features. Both molecular species are usually detected towards hot molecular cores \citep{bonfand17,fuente14,codella13,olmi96}. The rotational temperature and column density determined from the rotational diagram of the CH$_{3}$CN (T$_{rot} \sim 82$ K and 
N $\sim 10^{15}$ cm$^{-2}$) are in agreement with those determined towards several hot molecular cores (e.g. \citealt{araya05}). The mass of gas and the density obtained for the core agrees with the lowest values obtained in several molecular hot cores that were investigated using the 1.3 mm continuum and CH$_{3}$CN emission observed with the Submillimeter Array \citep{hern14}. A mass about 10 \msol~is the usually adopted lower limit for the gas mass
of a hot molecular core \citep{cesaroni05}. Comparing the mass of the core with the bolometric luminosity obtained towards this source,
about $4.5 \times 10^{4}$ L$_{\odot}$ (based on the luminosity of \citet{mottram11} corrected to the assumed distance of 6.2 kpc), and following \citet{cesaroni05}, we can conclude that we are observing a
``light'' hot molecular core, which implies that the mass of the embedded star is comparable to or greater than the gas mass.
\citet{hern14} found that most of the M$_{vir}$ values are greater than M$_{gas}$ in their hot molecular cores sample, as in our case, and they propose that this imbalance could be due to that the cores are not in dynamical equilibrium owing to complicated kinematics, and large rotating structures such as toroids and disks among other dynamical processes that will increase the line width and thus the virial mass. We conclude that we are indeed observing a ``light'' hot molecular core, which probably contains a rotating toroid/disk.

It is worth noting that the CH$_{3}$CN emission appears slightly elongated from southwest to northeast (see contours in Fig.\,\ref{comb} and the channel map at 101.0 \ks~in Fig.\,\ref{chann2}) in agreement with the inclination of the system as observed at near-IR wavelengths. Taking into account that this feature coincides with the dark lane observed at near-IR, it is also possible that part of the emission of CH$_{3}$CN traces a toroid as it was suggested in the hot molecular core G31.41+0.31 based on observations of CH$_{3}$CN J=12--11 \citep{cesa11}. Indeed, this molecular species is also detected towards protoplanetary disks \citep{loomis18}.

The morphology of the CH$_{3}$OCHO emission is more complex than the morphology of the CH$_{3}$CN. As  observed in the channel maps (Fig.\,\ref{chann1}), the CH$_{3}$OCHO emission extends along some filaments and concentrates
in knots and clumps similar to that found in the star-forming complex Sagittarius B2(N) \citep{sch19}. In that case the authors discarded that the 
CH$_{3}$OCHO filaments were due to outflows or explosive events like the one seen in Orion KL by \citet{bally17}, because such filaments are curved and 
bent. In G29-YSO we observe that at the first channel map (at 95 \ks~in Fig.\,\ref{chann1}) a straightforward and collimated structure extends towards the southeast in coincidence with the southern near-IR nebulosity. At the two last channel maps (at 105.5 and 107.0 \ks) two weaker filaments are observed: one elongated and pointing to the northern near-IR nebulosity, and the other one pointing to the M-shaped structure as seen in the near-IR emission. These structures mapped in the CH$_{3}$OCHO emission suggest expanding motions of molecular gas from the region of the dark lane where protostar(s) should be embedded, indicating outflow
activity in the region, in agreement with the scenario pictured from the near-IR data. The others CH$_{3}$OCHO structures present core morphology. In particular, towards the southeast, a core of this molecular species coincides with a core observed in the continuum emission at 1.3 mm. 

The CH$_{3}$OCHO emission shows the presence of molecular material mainly southwards, in coincidence with the 
region where the near-IR emission presents a sharper structure than in the northern case. This suggests that the jets and/or stellar winds are encountering a dense region, while towards the north they can flow more freely, generating the more extended features seen at near-IR. Additionally, it is worth noting that the 
distribution of the CH$_{3}$OCHO towards the northwest could explain the lack of near-IR emission in the region below the M-shaped structure described above (see Fig.\,\ref{comb}).

\subsection{Clump scales}

Figure\,\ref{figred} shows that YSO-G29 is located at the northwestern portion of a molecular clump belonging to an elongated cloud as mapped in the \8 J=3--2 line. The red-shifted molecular outflow traced by the \2 J=3--2 line (also detected in the \2 J=1--0 line by \citealt{li16}) extends towards
the northwest, the same direction as the northwestern nebulosity detected in the near-IR observations. As mentioned above, and in agreement with previous works, we did not detect any blue-shifted molecular outflow. One possible explanation is that the YSO is located in the furthest edge of the molecular clump along the line of sight, and thus, the red-shifted molecular outflow freely flow to the exterior of the clump (and we can observe it), while the blue-shifted counterpart goes into the clump avoiding or difficulting its detection. This is consistent with the less intense and less extended southern near-IR emission in comparison with the northern one and with the spatial distribution of the CH$_{3}$OCHO emission discussed above.

The molecular species observed with ASTE confirm the presence of high-density gas towards YSO-G29. The detection of HCO$^{+}$ is consistent with the presence of molecular outflows 
(e.g. \citealt{sanchez13,rawlings04}), and the HCN and HNC, indeed tracers of high-density gas, can be used to trace the evolutionary stages of star-forming regions \citep{grani14}. A statistical tendency of increasing HCN/HNC abundance ratio was found from startless clumps 
(values about unity) to UC \hii~regions (values about 9) by \citet{jin15}. If we consider that the HCN/HNC integrated line ratio may resembles 
the abundance ratio (integration line ratio variations are driven by similar changes in the abundance ratio; \citealt{hacar20}), according to \citet{jin15}, our value of 2.3 would correspond to an active infrared dark cloud core (aIRDCc). Following \citet{hacar20}, who show that the HCN/HNC integrated line ratio can be used to estimate the kinetic temperature (T$_{\rm K}$) of the gas, by using their Eq. 3, we obtain T$_{\rm K} \sim 23$ K towards G29-YSO. This value is in quite agreement with the dust temperature at the YSO position (T$_{dust} \sim 26$ K) obtained 
from the dust temperature map\footnote{http://www.astro.cardiff.ac.uk/research/ViaLactea/} which was derived from the PPMAP procedure done to the Hi-GAL maps in the wavelength range 70--500 $\mu$m \citep{marsh17}.  
Our T$_{\rm K}$ obtained from the HCN/HNC ratio also agrees with T$_{\rm K} \sim 21.7$ K estimated from the NH$_{3}$ emission \citep{wie12} towards the ATLASGAL cold high-mass clump G29.85-0.06 which lies at the southern portion of the molecular cloud displayed in Fig.\,\ref{figred}, suggesting that the whole molecular cloud, analyzed at clump scales, has a temperature of about 25 K. As \citet{hacar20}, who used the HCN and HNC J=1--0 line, we also find a very good correlation between the T$_{\rm K}$ derived from the HCN/HNC ratio (in our case using the J=4--3 line) and the T$_{dust}$ obtained from IR data.

The C$_{2}$H is observed towards several types of interstellar regions (e.g. \citealt{nagy15} and references therein), and particularly, it seems to be almost omnipresent along the different evolutionary stages of massive star formation \citep{beuther08}. As photodissociation of larger carbon-chain molecules and polycyclic aromatic hydrocarbons (PAHs) is one of its possible formation processes, it is known that C$_{2}$H is a good tracer of PDRs. \citet{beuther08} observed 
an increase in the C$_{2}$H line widths with the evolutionary stage of the star-forming regions, with values of 2.8 for IRDCs, 3.1 \ks~ for high-mass protostellar objects (HMPOs), and 5.5 \ks~for UC \hii~regions. The values obtained towards YSO-G29 are consistent with HMPOs, which is in quite agreement with the results obtained from the HCN and HNC observations described above. This is because, as \citet{jin15} state, the averaged HCN and HNC spectra of aIRDCc and HMPOs appear to have similar peak intensities and line widths.

\section{Concluding remarks}

A complete understanding of the processes involved in the formation of the stars requires detailed multi-wavelength studies of individual objects at different spatial scales.  
In this work, using near-IR and molecular lines data we perform a deep study of the young stellar object G29.862$-$0.0044 relating the core and clump scales. The near-IR emission shows a likely disk-jet system. The study of the molecular gas at both spatial scales suggests the presence of a hot molecular core. The gas temperature at core and clump scale is about 80~K and 25~K, respectively, showing the presence of an internal heating source.

YSO-G29 exhibits a conspicuous asymmetric morphology at both spatial scales. The molecular outflow at clump scale is monopolar with the red lobe located towards the north. The near-IR and the CH$_{3}$OCHO molecular emissions, which map the core scale, suggest a scenario where the jet has flowed more freely towards the north in agreement with the direction of the red lobe of the molecular outflow.  Interestingly, unlike other studies at near-IR bands towards other sources in which it was not detected the redshifted jet due to extinction effects, in this work it appears as the brighter one, showing that this asymmetry cannot be due to the extinction. We observe that at core scale the jet/outflow activity is markedly asymmetric but not monopolar. This can be due to a highly inhomogeneous medium, which has consequences at the larger spatial scales.

\section*{Acknowledgments}

We thank the anonymous referee for her/his fruitful comments.  
M.B.A. is a doctoral fellow of CONICET, Argentina.
S.P. and  M.O. are members of the {\sl Carrera del Investigador Cient\'\i fico} of CONICET, Argentina.
M.R. wishes to acknowledge support from CONICYT (CHILE) through FONDECYT grant No1190684 and partial support from CONICYT project Basal AFB-170002.
This paper makes use of the following ALMA data: ADS/JAO.ALMA\#2015.1.01312.S. ALMA is a partnership of ESO (representing its member states), NSF (USA) and NINS (Japan), together with NRC (Canada), MOST and ASIAA (Taiwan), and KASI (Republic of Korea), in cooperation with the Republic of Chile. The Joint ALMA Observatory is operated by ESO, AUI/NRAO and NAOJ. The ASTE project is led by Nobeyama Radio Observatory (NRO), a branch of National Astronomical Observatory of Japan (NAOJ), in collaboration with University of Chile, and Japanese institutes including University of Tokyo, Nagoya University, Osaka Prefecture University, Ibaraki University, Hokkaido University, and the Joetsu University of Education.

\bibliographystyle{aa}  
\bibliography{ref}
\IfFileExists{\jobname.bbl}{}
{\typeout{}
\typeout{****************************************************}
\typeout{****************************************************}
\typeout{** Please run "bibtex \jobname" to optain}
\typeout{** the bibliography and then re-run LaTeX}
\typeout{** twice to fix the references!}
\typeout{****************************************************}
\typeout{****************************************************}
\typeout{}
}

\label{lastpage}
\end{document}